\documentclass[a4paper,fleqn]{cas-sc}

\usepackage[numbers,sort&compress]{natbib}

\usepackage{graphicx}
\usepackage{xcolor}
\usepackage{url}
\usepackage{hyperref}
\usepackage{multirow}

\begin{document}

\let\WriteBookmarks\relax
\def\floatpagepagefraction{1}
\def\textpagefraction{.001}

\author[1]{Sofiya Makar}
\affiliation[1]{Cyber Science Lab, Canada Cyber Foundry, University of Guelph, Canada; Email: smakar@uoguelph.ca, adehghan@uoguelph.ca, ayazdine@uoguelph.ca}

\author[1]{Ali Dehghantanha}

\author[2]{Fattane Zarrinkalam}
\affiliation[2]{School of Engineering, University of Guelph, Ontario, Canada; Email: fzarrink@uoguelph.ca}

\author[3]{Gautam Srivastava}
\affiliation[3]{Department of Math and Computer Science, Brandon University, Manitoba, Canada; Email: srivastavag@brandonu.ca}

\author[1]{Abbas Yazdinejad}




\title{Systemization of Knowledge (SoK)- Cross Impact of Transfer Learning in Cybersecurity: Offensive, Defensive and Threat Intelligence Perspectives}

\begin{abstract}
Recent literature highlights a significant cross-impact between transfer learning and cybersecurity. Many studies have been conducted on using transfer learning to enhance security, leading to various applications in different cybersecurity tasks. However, previous research is focused on specific areas of cybersecurity. This paper presents a comprehensive survey of transfer learning applications in cybersecurity by covering a wide range of domains, identifying current trends, and shedding light on under-explored areas. The survey highlights the significance of transfer learning in addressing critical issues in cybersecurity, such as improving detection accuracy, reducing training time, handling data imbalance, and enhancing privacy preservation. Additional insights are provided on the common problems solved using transfer learning, such as the lack of labelled data, different data distributions, and privacy concerns. The paper identifies future research directions and challenges that require community attention, including the need for privacy-preserving models, automatic tools for knowledge transfer, metrics for measuring domain relatedness, and enhanced privacy preservation mechanisms. The insights and roadmap presented in this paper will guide researchers in further advancing transfer learning in cybersecurity, fostering the development of robust and efficient cybersecurity systems to counter emerging threats and protect sensitive information. To the best of our knowledge, this paper is the first of its kind to present a comprehensive taxonomy of all areas of cybersecurity that benefited from transfer learning and propose a detailed future roadmap to shape the possible research direction in this area.

\end{abstract}

\maketitle

\section{Introduction}

Transfer learning (TL) is a popular Machine Learning (ML) method to reuse a pre-trained model's knowledge in a new model, reducing training time. It is widely studied and applied in various fields such as mechanical engineering, medicine, molecular biology, Industrial Internet of Things (IIoT), robotics and cloud computing.

ML is the ability of a computer to learn from data without explicit programming. It utilizes various learning techniques such as multi-task, active, and online learning. One of these techniques is TL. Transfer learning aims to enhance conventional machine learning methods by leveraging knowledge acquired from one or multiple source tasks and applying it to enhance learning in a related target task \cite{torrey_shavlik}. Thus, TL greatly reduces training time and the data required for efficient training. The formal definition of TL is as follows. Given some observations from the source domain(s) and its corresponding task(s) and some observations from the target domain(s) and its corresponding task(s), TL applies the knowledge from the source domain(s) to increase the performance of the decision functions of the target domain(s). A domain $D$ is defined in Eq. \ref{eq1} as a set of inputs $X$ from the corresponding feature space $X$ and the marginal probability distribution $P(X)$.

\begin{equation}\label{eq1}
\centering
D=(X,P(X))
\end{equation}

A task $T$ is defined in Eq. \ref{eq2} as labels $Y$ from the corresponding label space $Y$ and the decision function $f$, which is learned by a model from data.

\begin{equation}\label{eq2}
\centering
T=(Y,f)
\end{equation}

Many models output the predicted conditional probability distributions of inputs. Therefore the decision function is usually defined as in Eq. \ref{eq3}. 

\begin{equation}\label{eq3}
\centering
f(X) = P(Y|X)
\end{equation}

\begin{figure}[htp]
    \centering
    \includegraphics[width=0.9\linewidth,keepaspectratio]{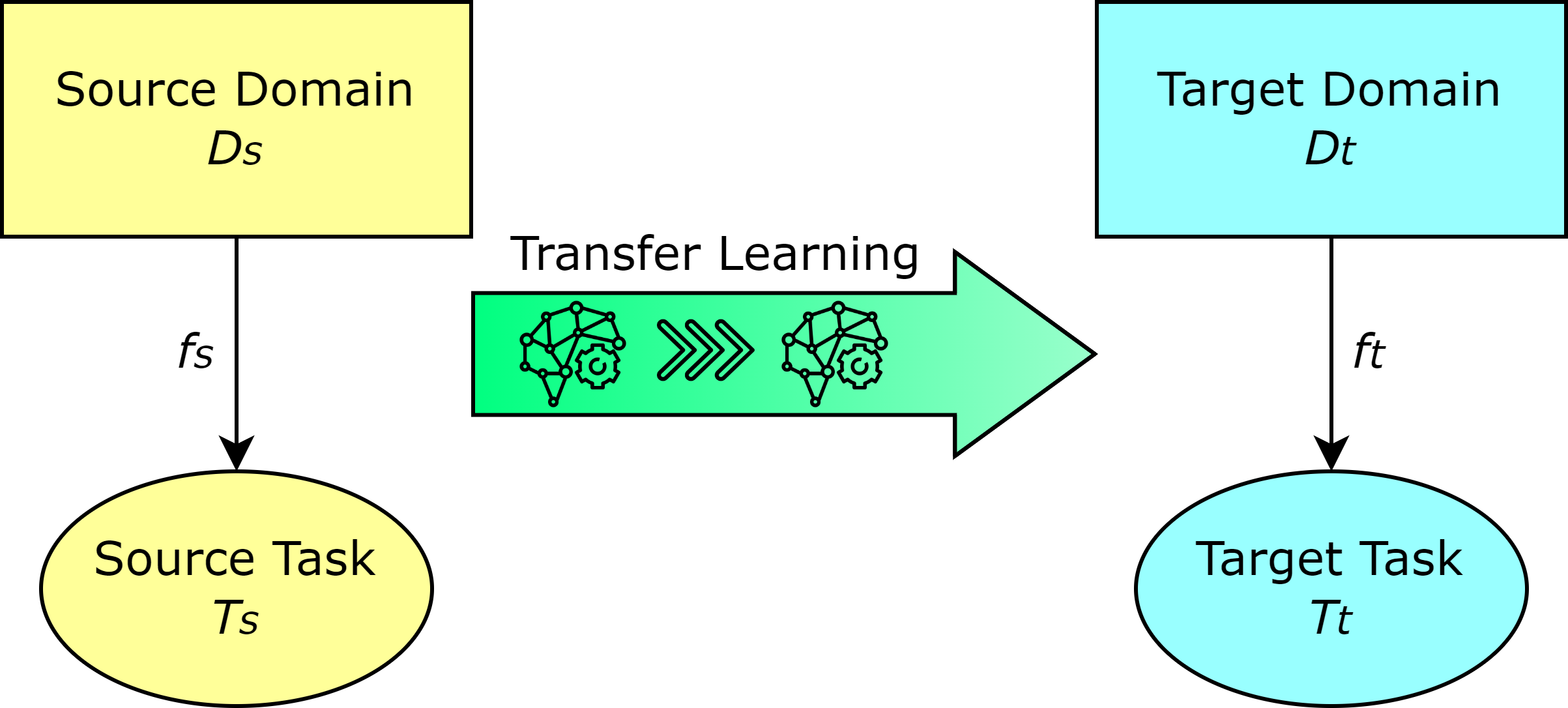}
    \caption{Definition of TL}
    \label{fig:definition}
\end{figure}

Figure \ref{fig:definition} provides a visualization of the TL process defined above. 

\begin{table*}[]
\tiny
\caption{Summary of Survey Papers}
\label{tab:survey}
\begin{tabular}{|p{2.35cm}|p{0.63cm}|p{6.56cm}|p{6.56cm}|}
\hline
\textbf{Survey Paper} & \textbf{Year} & \textbf{Pros} & \textbf{Cons} \\ \hline
Li \textit{et al.} \cite{survey_sec06} & 2022 & A detailed survey in one of the application areas of TL in security. The roadmap of possible future challenges is provided. Very recent work.  & Very narrow focus; the paper only considers the implementations of TL in the fault diagnosis area. \\ \hline
Zhao \textit{et al.} \cite{Trans-Sec-surv-Jour002} & 2021 & The authors constructed a taxonomy of the TL methods in the intelligent FD field and revealed the unseen challenges in this area. A practical analysis was performed. The authors published a test framework to address challenges and inspire future research. & Very narrow focus; the paper only considers the implementations of TL in the fault diagnosis area. The paper does not provide a conclusive future roadmap.  \\ \hline
Yin \textit{et al.} \cite{survey_sec01} & 2021 & A detailed survey on privacy-preserving federated learning. The paper proposed a taxonomy and a future roadmap. & The survey is not centred around TL; it only includes this approach as one of the methods to achieve privacy-preserving FL. Narrow focus.\\ \hline
Mothukuri \textit{et al.} \cite{survey_sec04} & 2021 & The work studies the privacy aspects of FL. Includes a future roadmap. & The survey is not centred around TL; federated transfer learning is one of the classes of FL. \\  \hline
Silva \textit{et al.} \cite{UCT-Jour001} & 2019 & The survey explores the TL methods developed to improve the performance of reinforcement learning agents. Includes a taxonomy. & Secure TL is not the primary focus of the survey. Does not provide a future roadmap.   \\  \hline
\end{tabular}
\end{table*}

The recent literature highlights a significant cross-impact between TL and cybersecurity \cite{a1,a2}. On the one hand, TL can be implemented to improve the performance of various cybersecurity tasks. On the other hand, security methods can be integrated with TL to improve the security of TL-based models.

A thorough survey of TL and cybersecurity can drive further research by highlighting current trends and under-explored areas. This paper first examines prior surveys, highlighting their limitations and motivating the need for a new survey. Next, we review the literature to answer the critical research questions:

\begin{itemize}
    \item \textbf{RQ1}: What are the main application and state-of-the-art of TL in cybersecurity (Section \ref{disc})?
    
    \item \textbf{RQ2}: What are the unexplored areas and possible future research directions for TL in cybersecurity (Section \ref{fut})?

    \item \textbf{RQ3}: What are the specific benefits of TL in the field of cybersecurity, and how do they contribute to enhancing the performance, efficiency, and resilience of cybersecurity systems (Section \ref{conc})?

\end{itemize}

To address these questions, we conduct a comprehensive survey of TL approaches applied in various cybersecurity fields. Such a detailed and specific survey can pave the way for further study in this field and greatly assist the researchers. The literature comes with several surveys on TL and its implementations in different security domains. The benefits and drawbacks of the selected survey papers are summarized in Table ~\ref{tab:survey} to compare with our work. As shown in the table, our paper is the first work focusing on all implementations of TL in the field of computer security. It presents the related taxonomies as well as a future roadmap. We will examine various studies in the field to answer research questions in the following sections.

We summarize our main contributions as follows:

\begin{itemize}
    \item This paper is a first-of-its-kind literature review paper on the implementations of TL in the computer security area. By examining a wide range of applications, it offers a holistic view of the potential of TL in addressing cybersecurity challenges. 
    \item  Building upon the analysis of existing research, this paper proposes a comprehensive taxonomy for TL in cybersecurity. The taxonomy categorizes TL applications based on the specific cybersecurity domains and subdomains. This taxonomy offers a structured framework for understanding and classifying TL approaches in cybersecurity.
    \item  In addition to the analysis and taxonomy, this paper provides additional insights into the benefits and challenges of TL in cybersecurity. It highlights the advantages of TL and identifies its challenges. These insights contribute to a deeper understanding of the implications and potential future directions of TL in cybersecurity.
    \item Lastly, this paper outlines future research directions and challenges that require the attention of the cybersecurity community by highlighting the unexplored areas of TL in cybersecurity. 
\end{itemize}

The remainder of this paper is structured as follows: Section \ref{meth} details the methodology used in this work. Section \ref{disc} reviews the state-of-the-art literature and develops the taxonomy of TL-cybersecurity cross-impact, Section \ref{fut} outlines the future roadmap, and Section \ref{conc} concludes the paper.


\section{Methodology}\label{meth}

We have conducted a systematic literature review to develop a comprehensive overview of the research field that studies the cross-impact of TL and computer security. 

We performed our search by combining various keywords and inputting the resulting query into the search engine of selected digital libraries. We choose the following platforms for computer science publications: IEEExplore, ACM Digital Library, SpringerLink, ScienceDirect, and Google Scholar.

We searched the publications based on the following keywords:

("transfer learning" AND ("malware" OR "security" OR "privacy" OR "attacks" OR "cybersecurity" OR "cybersecurity"))

The searches were conducted on multiple occasions during the period from October 2021 to January 2022. We identified 143 papers published during or before this period that met the keywords criteria. Next, we filtered the obtained publications based on the following selection criteria: the paper must have a focus both on TL and any field in cybersecurity, the paper must be written after 2018 and published in English, and the paper must not be a pre-print.

The research area of TL is relatively new, and only a limited number of publications are available for review, especially after filtering out all security-unrelated works. This fact stimulated us to relax most of the exclusion criteria. Out of 136 publications yielded by a primary search, we removed 65 by applying selection criteria, leaving us with 78 papers to review. Our search encompassed the following libraries: IEEE (39 papers), ACM (10 papers), ResearchGate (9 papers), MDPI (5 papers), Science Direct (4 papers), Springer (3 papers), and Elsveir (1 paper). Furthermore, we examined the publication year of the collected papers. Our analysis revealed the following distribution: 2016 (1 paper), 2017 (5 papers), 2018 (10 papers), 2019 (9 papers), 2020 (23 papers), 2021 (26 papers), 2022 (3 papers), and 2023 (1 paper).

\section{State-of-the-Art}\label{disc}

\begin{figure*}[htp]
    \centering
    \includegraphics[width=1\linewidth,keepaspectratio]{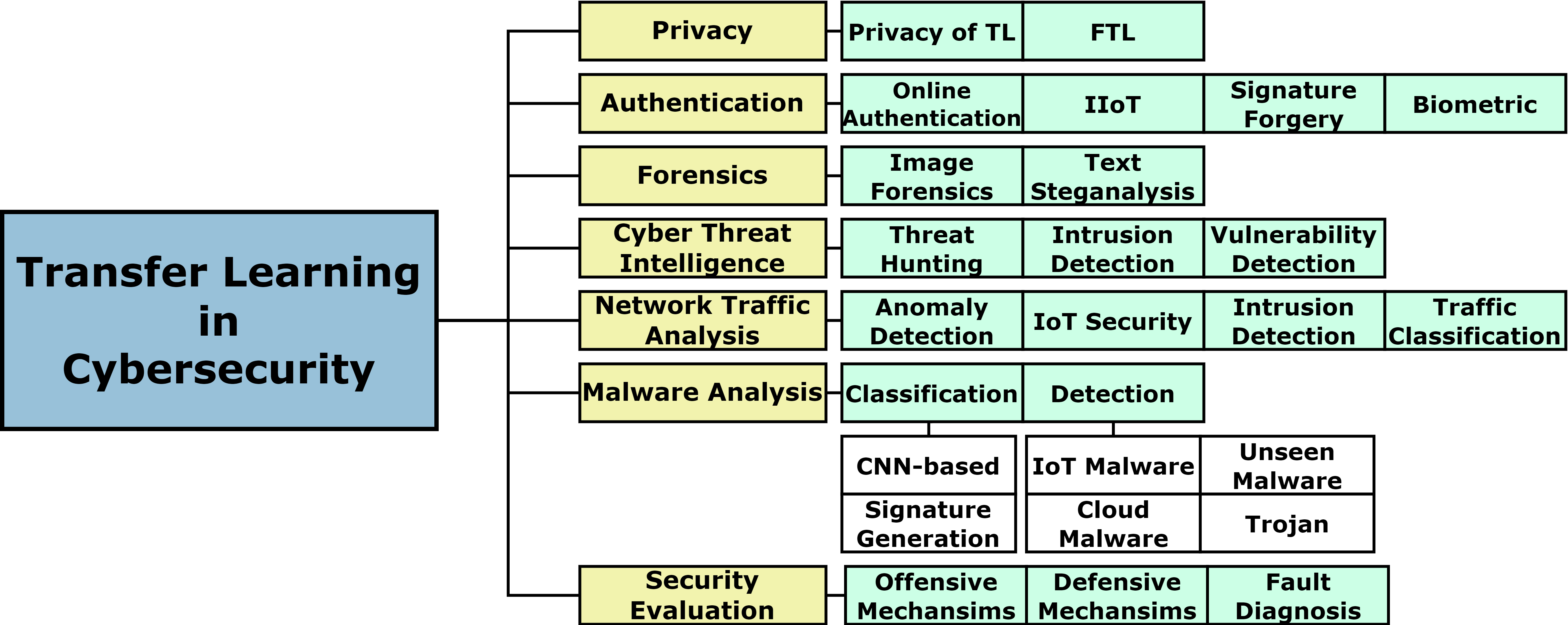}
    \caption{Taxonomy of the implementation of TL in cybersecurity}
    \label{fig:tax}
\end{figure*}

This section provides an in-depth exploration of the current advancements and practical implementations of TL in the field of cybersecurity. The literature has been systematically organized into seven primary application categories, forming the foundational layer of the proposed taxonomy (Fig. \ref{fig:tax}). The literature has been further segmented through rigorous analysis into subcategories, constituting the second layer of the taxonomy. Notably, a distinct emphasis on malware analysis is observed, prompting the identification of commonalities among the relevant studies and the subsequent development of a supplementary third layer within the taxonomy.

\subsection{Privacy}

Privacy is a paramount concern in today's highly digitalized world, and ensuring the protection of sensitive information is crucial for any organization. This section explores the applications of TL in privacy-related aspects of cybersecurity.

Neural networks (NN) pose privacy challenges, particularly in protecting training datasets that may contain sensitive information. Jin \textit{et al.} addressed this challenge by incorporating Fully Homomorphic Encryption (FHE) in the NN training framework \cite{Behr-New-Conf001}. FHE works by allowing the computation on the encrypted data without the key. The previous work on FHE is highly limited to trivial applications since FHE requires vast computational resources. The authors proposed a new framework that efficiently implements the Gradient Descent algorithm in the encrypted domain, supporting multi-class classification networks with double-precision float-point weights. By applying TL, the authors extended the functionality of their framework to be applied to complicated real-world problems. 


Hu \textit{et al.} identified the problem of privacy leakage in TL for recommendation systems, where existing research primarily focuses on improving target domain performance while neglecting the privacy of the source domain \cite{hu-yang-2020-privnet}. To address this issue, the authors proposed a privacy-aware neural representation called PrivNet. By simulating attacks during training and utilizing an adversarial game framework, PrivNet effectively protected the privacy of unseen users while enhancing the target performance. Experimental results demonstrated the successful disentanglement of knowledge beneficial for transfer without compromising privacy. 


\subsubsection{Federated Transfer Learning}
FTL is a technique that combines two important concepts: FL and TL. It aims to address the challenge of integrating and utilizing data from multiple organizations or domains while preserving privacy and improving the performance of machine learning models. In traditional TL, knowledge learned from a source domain is transferred to a target domain to improve the performance of the target model. However, FTL extends this concept by considering a scenario where data is distributed across multiple organizations or domains \cite{a3}. Each organization retains control over its data and does not share it directly with others due to privacy concerns and legal constraints. 

FTL was first introduced by Liu \textit{et al.} in 2018 \cite{Security-of-Trans-Conf006}. The authors aimed to improve the data integration process across different organizations for machine learning. They developed FTL to enhance statistical modelling in a data federation setting. The proposed framework requires minimal modifications to existing model structures and achieves the same level of accuracy as non-privacy-preserving TL. It offers flexibility and can be effectively adapted to various secure multiparty machine learning tasks. 

A year after, Sharma \textit{et al.} noticed that the previous implementation suffered from excessive computational overhead, rendering it impractical \cite{Security-of-Trans-Conf003}. To enhance efficiency and security, the authors incorporated Secret Sharing (SS) into the FTL model and extended it to handle malicious players who may deviate from the protocol. By utilizing the multi-party computation SPDZ protocol, their model achieved significant improvements in the runtime and communication cost compared to previous work, reducing the execution time from 35 seconds to 0.8 seconds (semi-honest case) and 1.4 seconds (malicious case) for 500 samples.  

In the meantime, Gao \textit{et al.} \cite{Trans-Sec-Conf004} considered a different way to improve FTL. The authors addressed the limitations of existing FL approaches in handling covariate shift and feature heterogeneity without compromising privacy. They propose a TL approach called Heterogeneous Federated Transfer Learning (HFTL) to address these challenges while preserving stringent privacy. The HFTL framework incorporates privacy-preserving multi-party learning using homomorphic encryption and secret-sharing techniques. Experimental results demonstrate the security, effectiveness, and scalability of HFTL on benchmark datasets, and its application to in-hospital mortality prediction from the MIMIC-III dataset, highlighting its significance in privacy-sensitive scenarios. 

Wang \textit{et al.} proposed the feature-based federated transfer learning (FbFTL) method as an innovative approach for improving the communication efficiency of federated learning via wireless links while preserving client privacy \cite{Privacy-New-1}. FbFTL reduces the uplink payload by over five orders of magnitude compared to existing approaches by uploading extracted features and outputs instead of parameter updates. The system design and learning algorithm are described, and a random shuffling scheme is analyzed for privacy preservation. Experimental results demonstrate the effectiveness of FbFTL in significantly reducing uplink payload and ensuring privacy preservation, making it a communication-efficient and privacy-preserving novel FTL scheme.

FTL has garnered considerable interest within the academic research community, finding applications across diverse security domains including FD, image steganalysis, network analysis, and traffic classification. The subsequent sections will provide comprehensive discussions on selected implementations of FTL in these domains \cite{a4}.

Overall, the ongoing advancements in privacy-preserving TL methodologies offer promising solutions for addressing privacy concerns and facilitating the development of secure and effective machine learning systems. However, further research is needed to enhance privacy preservation, improve model robustness, and explore the applicability of these techniques to various real-world scenarios.

\subsection{Authentication}

Authentication is pivotal in ensuring secure access and identity verification in various domains, from online services to physical premises \cite{a5}. With the increasing reliance on digital systems and the proliferation of sensitive data, robust and reliable authentication mechanisms have become paramount in mitigating unauthorized access and protecting user privacy. This section delves into the realm of authentication in the context of TL, exploring innovative approaches, challenges, and advancements that leverage TL techniques to enhance authentication systems. We will delve into different aspects of authentication, including TL for improving online authentication systems, securing authentication in the context of the IIoT, detecting signature forgery and enhancing human biometrics authentication.

Chen \textit{et al.} presented TL-PHA (TL-based physical-layer authentication), a novel physical-layer authentication scheme designed for fast online user authentication in latency-sensitive applications \cite{AI-Auth-Jour002-1}. TL-PHA utilizes the triple-pool network, a lightweight CNN architecture, for efficient online classification. Additionally, effective data augmentation methods are incorporated to enhance the training dataset. Experimental results, using both simulated channel data and real experiment data, demonstrate the superiority of TL-PHA in terms of authentication accuracy, detection rate, and training complexity compared to alternative approaches. The proposed TL-PHA scheme represents a significant advancement as the first physical-layer authentication scheme for latency-sensitive applications that employ TL.

Authentication plays a crucial role in ensuring secure and reliable communication within the IIoT ecosystem \cite{a6}. With the rapid growth of IIoT applications, ensuring the authenticity and integrity of user identities has become paramount. Wang \textit{et al.} addresses the data security and privacy challenges in IIoT applications by proposing a novel authentication mechanism based on TL-empowered Blockchain (ATLB) \cite{AI-Auth-Jour014}. The proposed ATLB combines blockchain technology for privacy preservation and TL for efficient user authentication. Different blockchains are introduced to counter collusion and Sybil attacks, while user authentication accuracy is enhanced through credit-based authentication mechanisms and a guiding network-based deep deterministic policy gradient algorithm. Additionally, TL is applied to reduce training time and enable trustworthy blockchains. Experimental results demonstrate that ATLB provides accurate authentications and achieves high throughput and low latency in various IIoT scenarios. Similarly, Arumugam \textit{et al.} discussed the challenges related to data security and privacy during the collection of real-time and automatic data from observing applications in IIoT \cite{cit17}.  The authors introduced an FTL model for authentication and privacy preservation using the novel supportive twin delayed DDPG (S-TD3) algorithm. The proposed approach utilizes the FTL blockchain to preserve privacy and security in industrial applications. The authentication process is based on user credit, and the S-TD3 algorithm trains local authentication models with high accuracy. TL is employed to reduce authentication model training time by transferring models from the local level to foreign users using the outer blockchains. The experimental results demonstrate accurate authentication for local and foreign users, as well as higher throughput and low latency in various IIoT scenarios.

Signature forgery poses a significant threat to the integrity and security of authentication systems. It involves the act of falsely imitating an individual's signature to gain unauthorized access or manipulate sensitive information. Detecting signature forgery is crucial in ensuring the reliability and authenticity of signatures used for verification and legal purposes. Manikantha \textit{et al.} conducted research focused on signature forgery detection using deep learning models and Siamese architecture \cite{cit16}. The paper presents a comparative study of various deep learning models using the Siamese architecture for detecting signature forgery. TL was adopted to implement base twin networks in the Siamese NN. The results demonstrate that the VGG16 model using Euclidean distance and Gaussian Naïve Bayes classifier achieves a maximum accuracy of 100\% on the CEDAR and Kaggle datasets (the smallest datasets). ResNet50 achieves the highest accuracy of 98.29\% for detecting forgery in Chinese signatures from the ICDAR 2011 SigComp dataset. Other models like MobileNetV2 and DenseNet121 also exhibit high accuracies for specific datasets.

Human biometrics, such as fingerprints, facial features, iris patterns, and voice recognition, offer a promising approach to authentication due to their unique and intrinsic characteristics. Li \textit{et al.} focused on improving face recognition performance by leveraging deep learning techniques and TL \cite{GYU-Conf005}. Traditional face recognition methods often struggle with uncontrolled environmental factors and limited training samples. The authors propose a deep CNN combined with TL and sparse representation to address these challenges. The proposed method, named T-DFE (TL-based Deep Feature Extraction), aims to overcome the limitations of traditional CNNs on small sample tasks while simplifying computational complexity. The approach involves training a CNN on augmented databases and applying TL to recognize faces with limited training samples. Experimental results using different databases demonstrate the superiority of the proposed method compared to traditional approaches. T-DFE achieves higher recognition rates in various scenarios, such as ORL, IMM, and AR databases, outperforming methods like Local Binary Pattern. On a similar note, Bonazza \textit{et al.} aimed to compare classical ML and TL approaches for low-cost real-time face authentication \cite{GYU-Conf006}. The authors highlight the importance of minimizing the size of biometric data in an access control context, allowing storage on remote personal media. To meet these constraints, the study focuses on lightweight versions of algorithms. The experiments compare various methods, including Random Forest, Support Vector Machines, and TL using MobileNet v1 and v2 for face authentication. The evaluation considers authentication accuracy, storage size, and training and prediction computation times. The authors found that TL allows for improved accuracy in face authentication tasks, but the resulting networks are more time-consuming and bulkier compared to classical machine learning methods. Random Forest and Support Vector Machines fulfill real-time constraints and can be stored on a remote card, making them suitable for the given constraints. Salem \textit{et al.} presented DeepZeroID, a privacy-preserving cloud-based biometric verification system that utilizes homomorphic encryption and TL \cite{Enc-New-Jour003}. The system addresses the vulnerability of storing sensitive biometric data on the cloud by encrypting the data and performing computations on the encrypted form. A pre-trained deep neural network is used as a feature extractor, eliminating the need to train on sensitive data and reducing the risk of information leakage. Experimental results show that DeepZeroID achieves a verification F1 score of 95.47\% when verifying the combined features of iris and fingerprint inputs with zero false positives. 
Salunke \textit{et al.} focused on the application of continuous user authentication using keystrokes and mouse movement behavioural patterns \cite{cit18}. Traditional methods of authentication, such as passwords and biometrics, only provide static authentication. Continuous authentication, on the other hand, verifies the user's identity on an ongoing basis by analyzing their unique behavioural patterns. The study addresses the challenge of gathering sufficient data to establish a user's behavioural pattern, which can delay the implementation of continuous authentication for new users. To overcome this issue, TL is employed with a feed-forward neural network model. TL allows the model to leverage knowledge from previous learning tasks and improve accuracy with less data. The results showed that the TL model achieved 9.76\% higher accuracy than the model trained without any previous learning. This indicates that TL can enhance accuracy even with limited data, which helps expedite the onboarding process for new users.

The advancements in technology, particularly in machine learning and deep learning \cite{a8}, have paved the way for innovative approaches to authentication, addressing the limitations of traditional methods \cite{a7}. TL emerged as a powerful technique for improving authentication accuracy with limited data and reducing time resources. The combination of TL and privacy-preserving techniques showcased the potential for developing secure and efficient authentication systems. By utilizing pre-trained neural networks as feature extractors and applying encryption methods, it was possible to achieve high accuracy in biometric verification while preserving user privacy.

\subsection{Forensics}

Digital forensics plays a critical role in investigating and analyzing digital evidence to uncover cybercrimes and support legal proceedings \cite{a22,a23}. With the proliferation of digital devices and the increasing complexity of cyber threats, the field of digital forensics faces new challenges in efficiently and accurately identifying and extracting relevant information from various types of data. By leveraging pre-existing models and knowledge, TL can address data scarcity and improve the accuracy and efficiency of forensic analysis tasks. In the context of digital forensics, TL finds applications in various areas, including text steganalysis and image forensics. 

Text steganalysis involves detecting hidden information within textual data, such as hidden messages or encrypted content. The use of Deep Neural Networks (DNNs) in text steganalysis has shown promising results; however, the increased complexity of these models leads to longer inference times, limiting their practicality. To address this issue, Peng \textit{et al.} proposed a text steganalysis method based on multi-stage TL that enhances both inference efficiency and detection performance \cite{Trans-Sec-Jour002}. Experimental results demonstrate that the proposed method outperforms existing DNNs-based steganalysis methods in terms of detection accuracy and inference efficiency. 

Image steganalysis focuses on identifying hidden information within image files, often used for covert communication or data concealment. In image steganalysis, traditional deep-learning models require a large, diverse, and high-quality dataset for training. TL addresses the limitations of dataset quality, variety, and quantity. Ozcan \textit{et al.} proposed a novel approach to increase the success rate and decrease the error rate in detecting stego and cover images  \cite{Trans-Sec-Conf007}. The authors compare two series of models trained with and without TL. The experiments are conducted on two advanced steganography algorithms, HUGO and WOW, with varying payload rates. The results demonstrate that the TL-applied model outperforms the normal trained model in detecting stego images. TL allows the model to learn the fundamentals of the dataset and adapt more quickly to the operations of steganography. It improves the success rate and detection performance of the model, particularly on challenging datasets with lower payload rates. Similarly, Yang \textit{et al.} introduced FedSteg, which employs FTL to train a secure and personalized distributed model for secure image steganalysis \cite{Trans-Sec-Jour003}. FedSteg addresses the challenges of aggregating scattered steganographic images and preventing the leakage of private data. It enables participants to collaborate and train a general model without sharing raw data, thus preserving user privacy. Through extensive experiments on state-of-the-art steganographic methods, FedSteg demonstrates improved performance compared to traditional non-federated steganalysis approaches. 


Overall, TL plays a significant role in advancing the capabilities of digital forensics by enabling the development of robust and accurate forensic models with reduced data requirements and computational overhead. It empowers forensic experts to address complex challenges in the analysis of digital multimedia content and contributes to the ongoing evolution of the field \cite{a9}.

\subsection{Cyber Threat Intelligence}

Cyber Threat Intelligence (CTI) is a process of locating, collecting, and analyzing data to gain knowledge about potential threats to a given system \cite{a10}. In the realm of CTI, TL has emerged as a powerful approach to enhance the effectiveness of various applications. TL offers a means to leverage knowledge and expertise obtained from existing models or domains and apply it to CTI tasks such as threat hunting, intrusion detection, analyzing the source code of public adversaries, and predicting software vulnerabilities \cite{a11}. 

The active search for potential threats in the network is called threat hunting. It is used to dig up the hidden adversaries that have penetrated the layer of passive defence in the organization. Activity Recognition (AR) plays a crucial role in threat hunting by enabling the identification of suspicious or malicious behaviours and activities within a system or network. Using pre-trained frameworks for AR models yields poor performance due to the significant difference between activity environments. Khan \textit{et al.} proposed UnTran a framework that utilizes TL to generate a common feature space for both the source and target domains, enabling the recognition of unseen activities with limited labelled data in the target domain \cite{Trans-Sec-Conf022}. The framework combines the knowledge from a pre-trained activity model in the source domain with activity models based on raw and deep features in the target domain. The authors evaluate the UnTran framework on three real-world datasets and demonstrate its effectiveness in recognizing both seen and unseen activities in the presence of limited labelled data and imbalanced class distributions. 

Organizations use different intrusion detection systems to detect suspicious activities in their environment \cite{a12, a13}.  ML methods have been proposed for intrusion detection in Internet-of-Things (IoT), but they often require long training times and need to learn new models from scratch when the environment changes. Yilmaz \textit{et al.} addressed these issues by introducing a TL-based algorithm for intrusion detection in IoT \cite{New-New-Jour002}. TL is employed in two settings: transferring knowledge to generate suitable intrusion algorithms for new devices and transferring knowledge to detect new types of attacks. The results show that the TL approach outperforms traditional learning in detecting new attacks in terms of accuracy. The approach achieves better performance in various attack cases in different scenarios. 

Controller Area Network (CAN) bus is a standard system for in-vehicle communications. However, the CAN bus is susceptible to network-level attacks, and new types of intrusion attacks are constantly being discovered. Developing an efficient deep neural network-based detection mechanism for these attacks can be challenging without a large amount of intrusion data \cite{a24}. To address this challenge, Tariq \textit{et al.} proposed CANTransfer, an intrusion detection method for the CAN bus that utilizes TL \cite{GYU-Conf011}. The authors train a Convolutional Long Short Term Memory (LSTM) based model using known intrusion data to detect new attacks. By applying one-shot learning, the model can adapt to detect new intrusions with a limited amount of new datasets. The article presents extensive experimentation with CAN datasets collected from two different vehicles, KIA Soul and Hyundai Sonata. The proposed CANTransfer method achieves a performance gain of 26.60\% over the best baseline model for detecting new intrusions. 

As a tool to aid in CTI, a framework for automatically identifying and analyzing the source code from public adversaries' forums was developed by Ampel \textit{et al.} \cite{Trans-Sec-Conf018}. The study focuses on analyzing source code snippets posted on hacker forums, which are often noisy and unlabeled. To address this challenge, the authors propose a deep TL framework called Deep Transfer Learning for Exploit Labeling (DTL-EL) to collect and categorize hacker forum exploit source code. The DTL-EL framework leverages the learned representation from professionally labelled exploits to generalize better to hacker forum exploits. It classifies collected hacker forum exploits into eight predefined categories for proactive CTI. The performance of DTL-EL is compared to other models in the hacker forum literature. The authors concluded that DTL-EL's transferred layers better generalize to hacker forum source code compared to other deep learning and classical learning approaches. 

Yin \textit{et al.} discussed the importance of predicting the exploitability of software vulnerabilities \cite{Sof-Jour004}. While vulnerability descriptions contain rich semantic information, the size of the vulnerability description corpus is often too small to train comprehensive Natural Language Processing models. To address this limitation, the paper proposes a framework called ExBERT for accurate exploitability prediction. ExBERT is an improved version of the BERT model that involves fine-tuning a pre-trained BERT model using a domain-specific corpus of vulnerability descriptions. The TL approach captures the semantic information in vulnerability descriptions more effectively than other feature extraction methods. The experimental results demonstrate that ExBERT achieves state-of-the-art performance in exploitability prediction, outperforming other approaches in accuracy, F1 score, precision, and recall. 

Detecting threats in non-English Dark Net Markets (DNM), particularly in countries like Russia and China, is crucial due to the significant presence of cybercriminals in these regions. It allows for better monitoring and understanding of cybersecurity risks originating from these countries and enables effective countermeasures. Ebrahimi \textit{et al.} discussed this problem and addressed the challenges posed by the language barrier and limited labelled data \cite{Sof-Conf005}. Previous approaches have used machine translation to overcome these challenges, but translation errors can degrade threat detection performance. To improve detection, the authors proposed a deep cross-lingual TL model that learns a shared Bidirectional LSTM. TL, specifically cross-lingual knowledge transfer, is employed to capture common hacker-specific representations across languages. Deep learning techniques, such as BiLSTMs, are utilized to capture temporal patterns and extract transferrable features.

In conclusion, TL has emerged as a valuable approach for proactive CTI, enabling organizations to stay ahead of evolving cyber threats and protect their infrastructure effectively.

\subsection{Network Traffic Analysis}

Network traffic refers to the data in transit within a network during a specific period \cite{a14}. The continuous monitoring of network traffic is essential to identify any anomalous behaviour, accurately classify encrypted data packets, and promptly detect potential intrusion attacks. These activities play a critical role in maintaining the integrity, confidentiality, and smooth flow of information between servers, thereby ensuring a secure network environment\cite{a15}.

TL has played a crucial role in enhancing the effectiveness of various techniques for network traffic analysis and anomaly detection. TL has been successfully implemented in models for detecting abnormal network traffic, identifying anomalies in distributed data, and classifying different types of network traffic. It has addressed challenges such as scarcity of labelled data, widely distributed data, and data privacy concerns \cite{a25}. TL has also been instrumental in detecting malicious traffic, distinguishing between normal and malicious activities, and detecting previously unseen network attacks. Furthermore, TL has been applied to traffic classification tasks, enabling the identification of different types of applications flowing through networks while maintaining user privacy. TL has proven effective in dealing with encryption technologies and successfully identifying encrypted traffic. 

Network traffic comes in vast quantities; therefore, it is infeasible to monitor each packet. However, traffic between specific servers is often repetitive and predictable. For this reason, security specialists and researchers focus on detecting only abnormal traffic (the anomaly), which can potentially contain malicious data. Yang \textit{et al.} proposed a model to detect anomalies in communication network traffic \cite{Trans-Sec-Ear-Jour002}. A TL approach was used to deal with the scarcity of labelled data for abnormal traffic. Yehezkel \textit{et al.} proposed normalizing autoencoder losses in various models for detecting network abnormality \cite{GYU-Conf009}. Usually, autoencoders are used to learn the representation of normal traffic behaviour. The autoencoder loss is then calculated to identify any abnormalities. The proposed approach normalizes the loss to make it applicable to different network settings. This model will allow transferring the loss vector to detect the abnormalities of other networks via TL. Zhao \textit{et al.} addressed an issue of data scarcity \cite{Sof-Conf008}. An FTL anomaly detection model was proposed to solve this problem. The experiments showed that the model outperformed other anomaly detection models when dealing with sparse data. Xiong \textit{et al.} discussed a  similar problem of widely distributed data in the network traffic and proposed a successful solution using deep TL \cite{Sof-Conf004}. Wang \textit{et al.} explored how to detect abnormal physical nodes in network slicing \cite{Trans-Sec-Jour005}. A TL approach was successfully implemented in the model by exploiting similarities between the nodes. Aburakhia \textit{et al.} developed a CNN image-based anomaly detection model for extracting deep features and performing the detection analysis on these features was introduced \cite{Trans-Sec-Conf010}. Since pre-trained agents contain broad, deep feature representation, TL is naturally used to utilize such knowledge. 

IoT is the network of interconnected physical devices varying from household "smart" devices to large-scale industrial equipment\cite{a16}. However, the components of IoT can be vulnerable to outside attacks. Therefore, monitoring the IoT network traffic for any suspicious activities is important. Tien \textit{et al.} introduced the anomaly detection model for classifying different device types in the IoT network traffic \cite{Sof-Jour021}. First, the model learns the significant feature of the device types, and then, by applying autoencoders as a method for TL, the obtained knowledge can be applied on other sites. Wang \textit{et al.} proposed the CNN-based anomaly detection model for industrial control systems traffic analysis \cite{Trans-Sec-Jour004}. To improve the performance of the detection system and make it capable of identifying previously unknown attacks, the authors embedded TL into the residual CNN.

Network Intrusion Detection System (NIDS) was developed to aid organizations in monitoring their environments for suspicious occurrences and potential compromises. Deep learning-based NIDS requires vast volumes of labelled data to be trained efficiently. Singla \textit{et al.} developed a TL-assisted NIDS to overcome a lack of labelled data \cite{Sof-Conf007}. The resulting model showed better performance in terms of identifying unseen attacks compared to the models trained from scratch while requiring less training data. 
Taghiyarrenani \textit{et al.} discussed the same challenge and proposed another TL-based method to distinguish between normal and malicious traffic in various network systems \cite{Sof-Conf010}. TL solution proved effective when compared to the baseline models. Zhao \textit{et al.} developed a TL approach for detecting previously unseen network attacks using the knowledge gained from the known attacks \cite{Sof-Jour018}. The results suggested that the TL-based model performs the best in detecting unseen attacks compared to baselines. 

Traffic classification is a process of identifying the types of applications flowing in a network. Network traffic classification requires a secure processing system to maintain users' privacy. To achieve this, Majeed \textit{et al.} introduced a novel FTL classification framework \cite{Security-of-Trans-Conf002}. The authors implemented cross-silo security protocols to accommodate the privacy-preserving learning model. Singh \textit{et al.} developed a classification model to detect malicious traffic from Darknet \cite{Sof-Jour005}. The authors implemented TL to utilize the knowledge from ten pre-trained CNN models and achieved a 96\% accuracy rate of classifications. Li \textit{et al.} proposed a model for detecting unseen malicious network traffic \cite{Trans-Sec-Conf017}. The authors incorporated the adaptation regularization TL to deal with the issue of novel malware samples. The model outperformed other traffic classification and intrusion detection counterparts. Another research was done to detect malware in the interconnected network community by Rong \textit{et al.} \cite{Chap004}. The TL approach was implemented to solve sparse data distributions in the training and testing datasets.

One of the biggest challenges of traffic classification and analysis task is the encryption technologies implemented to improve the user's security \cite{a17}. These technologies allow malicious agents to hide their activities and evade detection. Zhang \textit{et al.} developed a model that adopts a TL approach based on Efficientnet to identify encrypted traffic successfully \cite{Trans-Sec-Conf003}. The experiments showed that the proposed framework reaches 100\% accuracy and recall rates while requiring a small amount of training data. 

\subsection{Malware Analysis} 
Malicious software, commonly known as malware, constitutes a form of program meticulously crafted and unleashed by malevolent individuals to execute harmful operations, including but not limited to pilfering sensitive information, assuming system control, or encrypting files for extortion purposes. In an effort to comprehend the intricate dynamics of a given malware specimen, experts undertake the crucial task of classifying it within a specific malware family via the process of malware classification. These families exhibit shared attributes that enable the development of distinctive signatures for malware detection and attribution, consequently empowering us to combat malware with increased efficiency and efficacy. 

Malware classification and detection have benefited significantly from the application of TL techniques \cite{a17}. TL has been successfully implemented with various deep learning models, such as Xception, VGG16, VGG19, and ResNet-50, to achieve high accuracy rates in classifying malware samples. TL has also addressed the challenges of unbalanced malware families, limited labelled data, and the need for efficient detection of unknown malware. Furthermore, TL has been instrumental in identifying malicious domain names and improving the detection of malware in cloud services and IoT devices. It has also shown promise in trojan attack detection in hardware systems. 

\subsubsection{Malware Classification}
NN-based malware analysis is often performed by representing a malware sample as a grayscale image to be processed by one of the CNN agents. One such agent is Xception which was designed to deal with the overfitting problem, which is common to most CNNs. Lo \textit{et al.} implemented TL to classify malware by the Xception model pre-trained on the Keras ImageNet dataset \cite{Sof-Conf002}. The model achieved the highest training rate at that time.

Marastoni \textit{et al.} compared the performance of LSTM and CNN models in the malware classification task  \cite{Sof-Jour010}. The experiments on a famous malware dataset often used to evaluate the performance of malware classification models, MALIMG, has shown that both models achieve similar accuracy rates. However, the LSTM model takes twice as long to perform the classifications compared to the CNN model.

Pant \textit{et al.} proposed another image-based malware classification model \cite{GYU-Conf004}. The authors used TL to transfer knowledge from the famous VGG16 model to create a custom framework to achieve a high validation accuracy. VGG16 is a CNN model that won an ImageNet (object detection and image classification) competition in 2014. VGG16 consists of sixteen deep layers: five convolution layers, three fully connected layers, five max-pooling layers, and one dense layer. Pant \textit{et al.} have transferred the weights of the pre-trained VGG16 models and added another two dense layers and one pooling layer to perform malware classification. The VGG16-based model achieved an 88.4\% accuracy rate. Prima \textit{et al.} have also applied the VGG16 layers for the malware classification task \cite{Sof-Jour020}. Their model has achieved 98\% accuracy on the MALIMG dataset.

The unbalanced malware families affect the performance of CNN-based malware classification algorithms. Bouchaib \textit{et al.} proposed the use of SMOTE (Synthetic Minority Oversampling Technique) to balance the dataset of families \cite{GYU-Conf003}. The authors have implemented TL to take advantage of the knowledge from the previously trained VGG16 model, which yielded 98\% accuracy. El-Shafai \textit{et al.} used different CNN agents pre-trained on the ImageNet dataset to perform malware classifications \cite{Sof-Jour022}. The VGG16-based model has achieved a 99.97\% accuracy rate on the MALIMG dataset.

Karanja \textit{et al.} proposed an image-based IoT malware analysis framework using TL to adopt layers from the pre-trained VGG19 model, a famous 19-level deep CNN \cite{sof-Chap003}. The adopted last layer discriminated the images of malware into corresponding malware families. The accuracy rate on the IoTPoT dataset is 89.23\%.


Kumar \textit{et al.} adopted the layers of implemented another famous CNN agent, ResNet-50, to classify the IoT malware \cite{Sof-Jour014}. The model showed a 99.18\% accuracy rate on the MALIMG dataset.

It is crucial to analyze malware and generate its signature to detect, classify, and possibly even attribute future samples of the same or closely related malware. Nahmias \textit{et al.} introduced a novel malware signature generation method, TrustSign \cite{Sof-Jour008}. The model is entirely unsupervised and uses deep features transferred from a pre-trained VGG-19 framework.

\subsubsection{Malware Detection}


It is highly challenging to detect previously unknown malware as its samples are absent in the training datasets. Therefore, this topic has received special attention from the community. As the signature-based malware detection algorithms fail to detect previously unseen malware, Zhang \textit{et al.} proposed a novel portable executable files feature extractor combined with the k nearest neighbours algorithm via TL to detect new malware samples successfully \cite{Sof-Jour001}. Similarly, Fu \textit{et al.} introduced an LSTM model combined with a generative adversarial network to detect unseen malware for mobile devices \cite{Sof-Jour002}. In addition, the TL approach was implemented to create uniform feature spaces and deal with a lack of labelled data. The proposed models are designed for detecting various kinds of attacks and intrusions. Sameera \textit{et al.} proposed a TL-assisted model to compensate for the lack of labelled data in zero-day attack detection \cite{Sof-Jour007}. 

The TL techniques were implemented in the malicious domain names detection model by Rajalakshmi \textit{et al.} \cite{Sof-Conf011}. It is challenging to detect malware domains since they are randomly generated by domain-generated algorithms. Moreover, such detection models must be implemented on the fly to be helpful in real-world scenarios. The authors combined the state-of-the-arc CNN model with ML classifying algorithms via TL. The final model successfully performed detections and classifications of malware-generated domains. 

Alshehri developed a TL-assisted image classification network to avoid the bottlenecks (various analysis obstacles) in the existing malware detection methods \cite{Sof-Chap001}. The author had combined two rapidly developing methods, namely TL under deep neural networks and Android malware detection algorithms, to improve the existing detection frameworks.

Some research was done to detect malware attacks in cloud services. Sreelatha \textit{et al.} utilized TL techniques to apply the knowledge from source domains to identify any abnormal activity in the communications and improve the detection rates \cite{Trans-Sec-Conf008}. To protect the privacy of cloud tenants while using their data for cloud computing malware detection and classification, Gao \textit{et al.} proposed a TL-assisted classifier model. The experiments showed that TL improves the detection accuracy from 94.72\% to 96.9\% \cite{Sof-Jour013}. 

As it was mentioned before, IoT remains vulnerable to outside attacks \cite{a18}. The ability to quickly and efficiently detect IoT malware is crucial in ensuring the security of the vast range of tools, equipment, and devices connected to IoT. Because of the rapidly-evolving nature of IoT, the labelling process of IoT samples is highly time-consuming. Vu \textit{et al.} proposed an autoencoder-based deep TL model to detect IoT malware which does not require fully labelled data \cite{Sof-Jour009}. The results show that the model improves detection rates compared to baselines. Bots remain a considerable threat to IoT. Taheri \textit{et al.} proposed to transform network traffic data into images and train a CNN model to recognize bot activity \cite{Sof-Jour012}. The study showed that using TL improved the accuracy from 33.41\% up to 99.98\%.

Trojan attack detection in hardware systems has received particular attention in the community. Trojans are any type of malware intended to mislead the user regarding its true objective. A hardware Trojan is a malicious modification in the hardware circuit systems that affect the system's functionality. Sun \textit{et al.} proposed an electromagnetic side-channel signal analysis method for identifying trojan attacks \cite{Trans-Sec-Ear-Jour003}. A TL network is trained on time-frequency data to classify the attacks. The experiments showed an improvement compared to a standard trojan detection method. A TL-assisted improvement for traditional trojan hardware detection tools was developed by Faezi \textit{et al.} \cite{Trans-Sec-Conf016}. The proposed neural network model is pre-trained on the side-channel signals of the known trojans, and the obtained knowledge is used to report the behaviour of particular hardware chips.

\subsection{Security Evaluation}

Organizations regularly evaluate their security systems to determine how well they perform against new vulnerabilities \cite{a19,a20}. The use of TL has been explored in various aspects of security analysis and defence mechanisms. TL has shown promise in profiled side-channel attacks, key recovery, CAPTCHA attacks, and improving the efficiency and accuracy of cryptographic models. However, vulnerabilities in TL frameworks have also been identified, such as attacks on the TL feature extractor, weaknesses in the Softmax layer of CNN classifiers, and backdoor attacks. To mitigate these risks, researchers have proposed defence models, backdoor detection techniques, and encryption algorithms. TL has also been applied in fault diagnosis systems to enhance performance, address data privacy concerns, and handle domain shift challenges. 

The TL algorithm was exploited by Garg \textit{et al.} to perform profiled side-channel attacks with two-dimensional CNN models \cite{Attack-Using-Trans-Conf001}. The side-channel analysis can help attack the private keys by identifying the weak spots in cryptographical algorithms. Side-channelling analysis can be used as an efficient technique to compromise secret keys. Thapar \textit{et al.} proposed a novel deep-learning side-channelling analysis model for more successful key recovery \cite{Attack-Using-Trans-Conf002}. TL was implemented to attack devices belonging to different families. CAPTCHAs are widely used to differentiate human users from potentially malicious bots. Threat actors have been developing various techniques to attack and evade CAPTCHAs; however, most are too complex and time-consuming. Wang \textit{et al.} conducted a quick and straightforward attack on text-based CAPTCHAs using TL to train the model on a few real-world samples \cite{Attack-Using-Trans-Jour001}. A successful adversary attack on the TL feature extractor model was performed by Abdelkader \textit{et al.} \cite{Security-of-Trans-Conf001}. The attack lowered the accuracy of the CNN classifier by 40\%, proving the vulnerability of the current TL frameworks. 

In cryptography, keeping the private key secure is the highest priority task. Encrypting malicious traffic is a popular method used by adversaries to avoid detection by security models. Cui \textit{et al.} conducted experimental research on attacking the keys. TL was implemented to determine the initial weights of the neural network to reduce the training time and improve the model accuracy while dealing with a lesser amount of data \cite{Trans-Sec-Conf001}.

On the other hand, some researches focus on improving defence mechanisms. Zhang \textit{et al.} proposed a defence model against the white-box attack on the TL framework by fine-tuning the parameters \cite{GYU-Conf013}. In addition, the authors conducted a black-box attack on the target model. The possible danger of backdoor attacks was addressed by Wang \textit{et al.} \cite{Rec-Year-Jour003}. Due to the increasing popularity of TL, more examples of pre-trained (Teacher) models are available online, exposing them to backdoor attacks. The authors successfully attacked the learning system, which resulted in significant misclassification rates. Wu \textit{et al.} explored how to defend TL models from the above-mentioned misclassification attacks \cite{Security-of-Trans-Conf007}. The authors proposed a distilled differentiator to protect from both targeted and random attacks. The encryption algorithms can also be applied to ensure the better security of TL-based models. The idea to use a private key to ensure the better protection of TL systems from unauthorized access was presented by Pyone \textit{et al.} \cite{Security-of-Trans-Conf004}. TL allows for practical training of the large protected model utilizing small portions of the training dataset. 

FD is a process of determining which fault has occurred in a system. Because of the development of ML algorithms, we have observed a noticeable improvement in FD techniques. Chen \textit{et al.} addressed the lack of labelled data needed for successful FD and proposed to use TL to resolve this challenge \cite{New-New-Conf002}. The proposed model can perform efficient diagnoses even with a limited number of fault samples. Zhang \textit{et al.} addressed the challenges of data privacy and domain shift in collaborative machinery fault diagnosis \cite{Trans-Sec-Ear-Jour001}. The authors proposed an FTL method to enable joint model training while ensuring data privacy. The method involves a federal initialization stage to maintain consistent data structures in distributed feature extraction and a federated communication stage using deep adversarial learning. Additionally, a prediction consistency scheme is incorporated to enhance model robustness. Experimental results on real-world datasets demonstrate the promising potential of the proposed FTL method for practical industrial applications in fault diagnosis.

The abovementioned advancements demonstrate the potential of TL in enhancing security and improving fault diagnosis techniques in practical applications.

\section{Future Roadmap}\label{fut}

To pave the way for future research and advancements in TL in the field of cybersecurity, several insights and considerations have emerged from our analysis. Researchers interested in continuing this work are encouraged to broaden the scale of TL applications to include the newest research developments, as the interest in TL algorithms continues to increase \cite{a22,a21}. Exploring novel application areas and domains where TL can be effectively applied will foster the growth of TL in cybersecurity.

While significant progress has been made in preserving privacy using TL-based models, privacy preservation remains an active area of research interest. The possibility of reverse-engineering the transferred knowledge from the source domain and recreating the original labels is a notable concern. Future research should focus on developing robust privacy-preserving mechanisms and techniques to ensure that sensitive information remains secure during knowledge transfer. Addressing the challenge of protecting the privacy of source domain data and preventing information leakage will be crucial for the widespread adoption of TL in computer security.

Furthermore, TL has often been employed to adapt knowledge from one or more CNN layers in various classification problems. Researchers typically manually freeze and transfer these layers to align the pre-trained CNN model with their specific research problem. It is foreseeable that automatic and flexible tools will emerge to expedite the knowledge transfer process from popular CNN models. These tools will simplify the adaptation of pre-trained models to new domains, enhancing the efficiency and accessibility of TL in cybersecurity applications.

Choosing a source domain that is closely related to the desired target domain is crucial for successful TL. Negative transfer, where the performance of the final model is diminished after employing TL, can occur if the two domains are not closely aligned. To address this challenge, researchers need a metric to measure the degree of relatedness between domains. Developing such metrics will be critical for guiding the selection of appropriate source domains and ensuring positive knowledge transfer. Resolving this issue will have a significant impact on the effectiveness and reliability of TL across all its application fields, including computer security.

Further advancements in TL algorithms are needed to improve their robustness and effectiveness in cybersecurity applications. Developing novel TL techniques that can handle more complex scenarios, such as handling adversarial attacks or incorporating domain-specific constraints, will be crucial. Exploring hybrid approaches that combine TL with other machine learning techniques, such as reinforcement learning or generative models, could also lead to improved performance and adaptability.

Finally, enhancing the explainability and interpretability of TL models in cybersecurity is of utmost importance. Developing methodologies to interpret the transferred knowledge and understand the decision-making process of TL models will improve their trustworthiness and facilitate their adoption in real-world applications. Researchers should focus on designing interpretable TL architectures and developing post hoc explainability techniques to shed light on the knowledge transfer process.

\section{Discussion and Conclusion}\label{conc}

This survey paper has been conducted in response to a need identified via an exhaustive search for a comprehensive review of the convergence of TL and computer security. We have analyzed seven key sections that highlight the diverse applications and potential of TL in addressing various challenges and concerns within the cybersecurity domain. Through this qualitative analysis of existing research, we have gained valuable insights into the advancements, methodologies, and limitations of TL in different cybersecurity contexts.

Moreover, this paper has provided additional insights into the methods and techniques employed in TL-based cybersecurity research. Deep learning models, such as CNNs and RNNs, have been extensively utilized, showcasing their capability to capture complex patterns and features in diverse cybersecurity datasets. Techniques like FTL have emerged to address privacy concerns while integrating data from multiple organizations or domains. 

Based on our research, we can conclude that TL algorithms have been successfully applied in a vast range of computer security fields. Among those, malware analysis was the most common application of TL. More specifically, TL was frequently used to transfer the knowledge from some layers of a pre-trained CNN model into a new image-based malware classification or detection model.

In addition, this paper has identified the most common problems that have been successfully addressed using TL in cybersecurity. These problems include the lack of labelled data, challenges arising from different data distributions, and privacy preservation concerns.

The scarcity of labelled data is a well-known challenge in the field of AI and cybersecurity. TL offers a solution by leveraging pre-existing models and knowledge gained from other domains or tasks. By transferring knowledge from these pre-trained models, the reliance on large amounts of labelled data is significantly reduced, enabling the development of robust cybersecurity systems with limited available data.

Another challenge arises from the presence of different data distributions in cybersecurity datasets. TL provides a means to bridge the gap between different domains or environments by leveraging knowledge obtained from one domain and applying it to another. By transferring the knowledge and learned features from a source domain to a target domain, TL allows the model to adapt and perform effectively in the target domain despite distributional differences.

Moreover, privacy preservation is a critical concern when dealing with sensitive data in cybersecurity. TL offers a way to address this challenge by utilizing pre-trained models that have already learned relevant features and patterns. By applying TL, the need to access and utilize the original sensitive dataset is eliminated, reducing privacy risks associated with data exposure.

Furthermore, the use of TL can also overcome the issue of inefficient training times. By leveraging pre-trained models, TL significantly reduces the time required to train a model from scratch. The pre-existing knowledge and learned features are transferred to the current model, allowing for faster convergence and more efficient training.

\bibliographystyle{unsrtnat} 
\bibliography{References}
\end{document}